\documentclass[runningheads]{llncs}
\usepackage{graphicx}
\usepackage{caption}
\usepackage{subcaption} 
\usepackage{booktabs} 
\usepackage[ruled,vlined]{algorithm2e}
\SetKwInput{KwInput}{Input}                
\SetKwInput{KwOutput}{Output}              
\usepackage[noend]{algorithmic}
\usepackage{overpic}
\newcommand{\comment}[1]{}

\usepackage{xcolor}

\newcommand{\m}[1]{\textcolor{red}{\textbf{#1}}}
\newcommand{\mah}[1]{\textcolor{blue}{\textbf{#1}}}

\usepackage{float}

\begin{document}
\title{Human-Like Summaries from Heterogeneous and Time-Windowed Software Development Artefacts}
\titlerunning{Human-Like Summaries from Software Development Artefacts}
%
\author{Mahfouth Alghamdi, 
Christoph Treude, 
Markus Wagner}
%
\authorrunning{M. Alghamdi et al.}
%
\institute{School of Computer Science, University of Adelaide, Adelaide, Australia
\email{mahfouth.a.alghamdi@adelaide.edu.au}\\  \email{christoph.treude@adelaide.edu.au} \\
\email{markus.wagner@adelaide.edu.au}}

\maketitle              
\begin{abstract}
Automatic text summarisation has drawn considerable interest in the area of software engineering. It is challenging to summarise the activities related to a software project, (1) because of the volume and heterogeneity of involved software artefacts, and (2) because it is unclear what information a developer seeks in such a multi-document summary.

We present the first framework for summarising multi-document software artefacts containing heterogeneous data within a given time frame. 
To produce human-like summaries, we employ a range of iterative heuristics to minimise the cosine-similarity between texts and  high-dimensional feature vectors. 
A first study shows that users find the automatically generated summaries the most useful when they are generated using word similarity and based on the eight most relevant software artefacts.


\end{abstract}
%
%
%

\section{Introduction and Motivation}

\begin{figure}[t]
    \centering\vspace{-2mm}
\includegraphics[width=\textwidth]{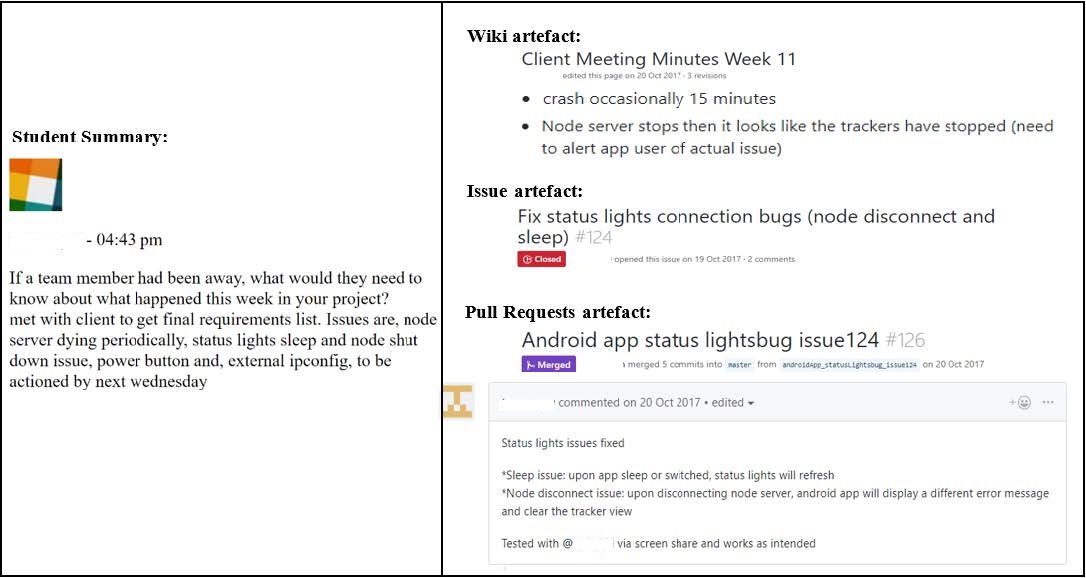}%
\vspace{-2mm}
\caption{An example of an anonymised student summary (left) linked to the content of related software artefacts (right).}
    \label{fig:studentSummaryExample}
    \vspace{-3mm}
\end{figure}

Modern-day rapid software development produces large amounts of data, e.g., GitHub~\cite{gitHub2020} now hosts more than 100 million repositories, with over 87 million pull requests merged in the last year, making it the largest source code hosting service in the world. The corresponding software development involves a lot of communication: developers create many types of software artefacts -- such as pull requests, commits, and issues -- 
and the amount 
can be overwhelming. For example, the Node\footnote{https://github.com/nodejs/node, accessed on February 2, 2020} project contains more than 11k issues, more than 20k pull requests, and over 29k commits. It also contains other software artefacts, such as wiki entries and readme files created by the developers during the project development life-cycle. In addition, these artefacts are frequently updated. For instance, in the week from January 1 to January 7, 2020, developers created 17 new issues, closed 12 issues and submitted 82 commits. 
Let us now consider two scenarios: (1) a developer has been on holidays during this period and would like to be updated, and (2) a new developer joins the team after this period and would like to know what has happened recently. In both cases, going through the artefacts and collecting the most useful information from them can be tedious and time-consuming. It is scenarios like these that we are targeting in our study, as solutions to these can ultimately increase the productivity of software developers and reduce information overload~\cite{treude2015summarizing}. 
To make these scenarios more tangible,  Figure~\ref{fig:studentSummaryExample} shows a summary written by a student software developer, as well as the various artefacts that contain parts of the information conveyed in this manually written summary. 

\newcommand\crule[3][black]{\textcolor{#1}{\rule{#2}{#3}}}

To offer solutions in such cases, we employ a combination of methods from 
Data-Driven Search-Based Software Engineering (DSE)~\cite{nair2018dse}. 
DSE combines insights from Mining Software Repositories (MSR) and Search-based Software Engineering (SBSE). While MSR formulates software engineering problems as data mining problems, SBSE reformulates SE problems as optimisation problems and use meta-heuristic algorithms to solve them. Both MSR and SBSE share the common goal of providing insights to improve software engineering. 
In this present paper, we suggest to improve software engineering -- in particular the creation of software development activities -- by mining the created artefacts for summaries. 

In recent years, several approaches have been developed to summarise software artefacts. Rastkar et al.~\cite{rastkar2014automatic} summarised bug reports using a supervised machine learning method. Rigby et al.~\cite{rigby2013discovering} summarised code elements from Stack Overflow using four classifiers based on context, location, text type, tf-idf, 
and element type. Furthermore, Nazar et al.~\cite{nazar2016source} and Ying et al.~\cite{ying2013code} utilised naive Bayes and support vector machine classifiers to generate summaries of code fragments taken from the official Eclipse FAQ. 
Interestingly, these techniques have mostly focused on summarising a single type of artefact, and 
they have not taken into consideration the production of summaries of content in a given time frame. 

To address these issues, we present a first framework to create multi-document summaries from heterogeneous software artefacts within a given time frame. 
In particular, we aim at an extractive approach, which generates a new summary from the relevant documents without creating new sentences~\cite{NenkovaMcKeown2011,VermaHari}.



The remainder of this paper is structured as follows. First, we 
describe the creation of the necessary gold-standard based on 503 human-written summaries 
in Section~\ref{sec:human}. 
In Section~\ref{sec:generatingsummaries}, we define the problem of summary-generation as an optimisation problem based on cosine-similarity and on 26 text-based metrics. 
In Sections~\ref{sec:experiments} and~\ref{sec:study}, we report on the results of our computational study and expert annotation of the results. We then discuss threats to validity in Section~\ref{sec:threats} and outline future work in Section~\ref{sec:conclusions}. 
\comment{
\paragraph{Background on automatic text summarisation.}
Due to the large amount of information on the internet, it has become difficult to simply skim all relevant source documents to extract pertinent information. Automatic text summarisation has attracted substantial interest for its potential to provide relevant information in less time by enabling the creation of short summaries of a single document or a collection of documents~\cite{luhn1958automatic}.

There are two strategies to automatically produce a summary: abstractive and  extractive~\cite{NenkovaMcKeown2011}. The abstractive approach aims at generating a new summary from the relevant documents after understanding the whole text. 
On the other hand, the extractive approach aimed at extracting a subset of the sentences from the relevant documents using linguistic and statistical properties of the text, such as sentence length and title similarity, cue phrases, and word frequency~\cite{VermaHari}.

Automatic text summarisation can be designed in terms of number of source documents in two ways: single document and multiple documents. A multi-document summary can be used to concisely describe the information contained in a set of documents to help the users understand different related aspects in these documents~\cite{alguliev2013multiple}. However, multi-document summarisation poses a challenge due the heterogeneity of the source documents~\cite{zopf-etal-2016-next}. Source documents containing heterogeneous data are expected to use different wording and topics, which can make it difficult for such a summariser tool to identify which sentence from which document should be selected to generate a summary.  

\m{I'm not happy with the position of this ``background'' in the RQ section}
}

\comment{
\textbf{RQ1} How does the content of students development activities produced on a weekly basis relate to artefacts in the project repository?

We are interested in 15 types of software artefacts, and the goal of our work is to find the software artefacts that contributed in the formation of the students' summaries by matching the sentences in these artefacts with the sentences found in the students' summaries. \mah{we can omit this coming sentence}We achieve this by utilising five heuristic algorithms, and we used cosine similarity in our optimisation algorithms
as the fitness function to guide the search toward the target vectors. \m{mixed up messages???}

\textbf{RQ1.1} How frequently are software artefacts referred to in the students' summaries? \m{why is this RQ1.1?}

Since a development activities in the students' summaries can be described using one or more artefacts, we are interested in the amount of the contribution of these artefacts to the generated summaries, hence we determine the common types of these artefacts as an input data.  

\textbf{RQ2} What are the upper and lower performance bounds for our optimisation algorithms? 

As this is the first study and the performance of the algorithms is unclear, we establish an upper bound using a brute force approach.
Additionally, the lower bound performance is established using random search. 

\m{careful: you are already answering the RQs here... I'm not sure this is common (Christoph?)}

Finally, we evaluate the automatically generated summaries with experts using Likert scale: 

\textbf{RQ3} How well are the summaries generated using our heuristic algorithms?
\m{text missing -- at least the other RQs have some text following}
}
\section{Human-written Summaries -- Creation of A Gold Standard}
\label{sec:human}

To better understand what human-written summaries of time-windowed software development artefacts look like, it has been necessary to create our own \emph{gold standard}. 
The basis of our gold standard is formed by a total of 503 summaries that were produced (mostly) on a weekly basis by 50 students over 14 weeks and for 14 (university-internal) GitHub projects. The students were working in teams of three or four on their capstone projects with clients from local industry toward a Bachelor degree (43 students in total) or with clients from academia toward their Masters degree (7 students in total). To ensure the usefulness of the students' summaries, each of the students' summaries was assessed as part of the student assessments during the particular semester. The summaries were anonymised before conducting this work to ensure confidentiality and anonymity of the students. 


We make use of these summaries to understand the general properties of human-written summaries, such as the summaries' typical length and the amount of information these summaries may contain. 
Additionally, the students' summaries can provide us with an understanding of the common types of artefacts related to development activities mentioned in the students' summaries and to help us identify which sentences from which artefacts should be selected for our extractive summarisation approach. 

To automatically collect the summaries, we used a Slack bot that asked the students to write summaries on a weekly basis about their projects development activity (see again Figure~\ref{fig:studentSummaryExample}). The written summaries were automatically recorded and collected in response to the question: \textit{If a team member had been away, what would they need to know about what happened this week in your project?} 

We show an example in Figure~\ref{fig:studentSummaryExample}: the question and the student's summary are on the left, and the relevant artefacts on the right. Note that the students only provide the summary, i.e., they do not provide a list of the relevant artefacts.



\begin{table}[t]
\caption{Number of artefacts contained in the 14 GitHub projects, resulting in 56,152 sentences.}
\centering
\label{tab:dataset}
\small{
\begin{tabular}{lrr}
\toprule
 \textbf{Type} & \textbf{Number} & \textbf{Sentences}\\
\midrule
Issue titles (IT) &1,885 & 1,885\\
Issue bodies (IB) &1,885 &5,650\\
Issue body comments (IBC) &3,280 &8,754\\
Pull requests titles (PRT) &1,103 & 1,103\\
Pull requests bodies (PRB) &1,103 &5,176\\
Pull requests body comments (PRBC) &897 &1,811\\
Pull requests reviews (PRRv) &2,019 &2,762\\
Pull requests reviews' comments (PRRvC)&1,286 &1,737\\
Commit messages (CM) &4,562 &7,856\\
Commit comments (CMC) &30 &55\\
Milestone titles (MT) &103 &103\\
Milestone description (MD) &103 &142\\
Readme files (RMe) &14 &2,678\\
Wiki files (Wiki) &492 &16,436\\
Releases (Rel) &1 &4\\
\bottomrule
\end{tabular}
}
\end{table}


We considered the following 15 types of textual artefacts in the GitHub repositories: issues (titles, bodies, and comments), pull requests (titles, bodies, comments, reviews, and reviews’ comments), commits (messages and comments), milestones (titles and descriptions), releases, wiki entries, and readme files. 
It is worth noting that, after manually inspecting the students' summaries, there was no evidence that any of these summaries contained a reference to a particular source code file.

The documents to be summarised as well as the summaries needed to first undergo various pre-processing steps -- including sentence splitting, stop-word removal, tokenisation, and stemming -- to reduce noise in the data. 
Also, we remove source code blocks from the software artefacts due to lack of evidence of student summaries citing code from actual files. 
Table~\ref{tab:dataset} lists the total amount of artefacts per type in our gold standard, as well as the total number of extracted sentences for each type.

\section{Methodology}
\label{sec:generatingsummaries}
In our approach, we intend to extract text from a set of heterogeneous software artefacts so that the resulting summaries are similar in style to those found in gold-standard summaries. 
In the following, we introduce two ways of measuring similarity, we revisit the definition of cosine similarity, and we define the iterative search heuristics used later on. 

\subsection{Generating Summaries Based on Word-Similarity and Feature Vector Similarity}

Historically, the selection of important sentences for inclusion in a summary is based on various features represented in the sentences such as sentence position~\cite{baxendale1958machine}, sentence length and title similarity~\cite{kupiec1995trainable}, sentence centrality~\cite{erkan2004lexrank}, and word frequency~\cite{luhn1958automatic}. Determining these features in the selection of important sentences is not simple and depends largely on the type of documents to be summarised~\cite{torres2014automatic}.

We consider two ways of characterising sentences: (1) based on the similarity of words, and (2) based on the similarity of feature vectors. In both cases, the goal is to select sentences from the collection of software artefacts so that the characteristics of the resulting summary are close to the characteristics of a target.

First, word similarity between two texts is defined by the number of times a term occurs in both texts, after the aforementioned stemming and removal of stop words. To achieve this, we use a vector-based representation, where each element denotes the number of times a particular word has occurred in the sentence. 

\begin{table}[t] 
\caption{Features used to represent each of the sentences.}\label{tab:Featurs}{\small
\begin{minipage}{0.48\textwidth}
\begin{tabular}{llcc}
\toprule
\textbf{No.} & \textbf{Feature} \\
\midrule
F1. &Word count\\
F2. &Chars count including spaces\\
F3. &Chars without spaces \\
F4. &No. of syllables in a word \\
F5. &Sentence length \\
F6. &Unique words \\
F7. &Avg. word length (chars) \\
F8. &Avg. sentence Length (words) \\
F9. &No. of monosyllabic words \\
F10. &No. of polysyllabic words \\
F11. &Syllables per word \\
F12. &Difficult words \\
F13. &No. of short words ($\leq3$ chars) \\
\bottomrule
\end{tabular}
\end{minipage} 
\begin{minipage}{0.51\textwidth}
\scalebox{1}{
\begin{tabular}{llcc}
\toprule
\textbf{No.} & \textbf{Feature} \\ 
\midrule
F14. &No. of long words ($>=7$ chars) \\
F15. &Longest sentence (chars) \\
F16. &Longest words (chars) \\
F17. &Longest words by number of syllables \\
F18. &Estimated reading time \\
F19. &Estimated speaking time \\
F20. &Dale-Chall readability index \\
F21. &Automated readability index \\
F22. &Coleman-Liau index \\
F23. &Flesch reading ease score \\
F24. &Flesch-Kincaid grade level \\
F25. &Gunning fog index \\
F26. &Shannon entropy\\
\bottomrule
\end{tabular}}
\end{minipage}
}
\end{table}

Second, as an alternative to the word-similarity and for situations where we a reference text is unavailable, we consider a total of 26 text-based features of sentences, which aim at capturing different aspects of readability metrics, information-theoretic entropy and other lexical features (see Table~\ref{tab:Featurs}). Each sentence is represented as a 26-dimensional vector of the feature values. For an initial characterisation of this high-dimensional dataset, we refer the interested reader to~\cite{Alghamdi2019giworkshop}.

\subsection{Cosine Similarity}

The most popular similarity measure used in the field of text summarisation is cosine similarity~\cite{manning2008introduction} as it has advantageous properties for high dimensional data~\cite{sohangir2017improved}. 

To measure the cosine similarity between two sentences $x$ and $y$ -- respectively their representation as a vector of word counts or the 26-dimensional representation -- we first normalise the respective feature values (each independently) based on the observed minimum and maximum values,
%
and then calculate the cosine similarity: 
\begin{equation}
\label{eqe:cosine_similarity}
\cos ({\bf x},{\bf y})= {{\bf x} {\bf y} \over \|{\bf x}\| \|{\bf y}\|} = \frac{ \sum_{i=1}^{n}{{\bf x}_i{\bf y}_i} }{ \sqrt{\sum_{i=1}^{n}{({\bf x}_i)^2}} \sqrt{\sum_{i=1}^{n}{({\bf y}_i)^2}} }
\end{equation}

We employ the cosine similarity in our optimisation algorithms as the fitness function to guide the search toward summaries that are close to a target vector.
\subsection{Algorithmic Approaches}

Extractive multi-document summarisation can be seen as an optimisation problem where the source documents form a collection of sentences, and the task is to select an optimal subset of the sentences under a length constraint~\cite{peyrard2016general}. In this study, we aim to generate summaries with up to five sentences as this is approximately the length of the summaries that the students have written.

We now present our optimisation algorithms to automatically produce summaries from heterogeneous artefacts for a given time frame. We utilise five algorithms, 
and we also create summaries at random to estimate a lower performance bound. We use the aforementioned cosine similarity as the scoring function, which computes either the word-similarity or the feature-similarity with respect to a given target. 
In our case, the targets are the summaries in the gold standard. 
By doing so, we aim at capturing the developers' activities found in the software artefacts that were created or updated in the given time frame and that are cited in the gold-standard summaries to generate human-like summaries.

Our first approach is a brute force algorithm, which exhaustively evaluates all subsets of up to a given target size.
We will use this as a performance reference, because 
we do not know a-priori what good cosine-similarity values are.


\begin{algorithm}[t]
{\small
\caption{Greedy algorithm}
\label{alg:greedy}
\KwInput{ 
    $AS$ - artefacts' sentences, $SS$ - student summary, 
    and
    $TLGS$ - targeted length of the generated summary.
}
\KwOutput{\textit{GS} – generated summary}
\begin{algorithmic}
\STATE $GS\gets $\O
\WHILE{($len(GS) \leq TLGS$)}
\STATE $K \gets$\O \{$K$: unused sentences in $AS$\}
\STATE $K_{best} \gets $\O \{best single sentence to add in this iteration\}
\FORALL{($K_{i} \in K$)} 
\IF{$cosSimilarity(GS + K_{i},SS) \geq cosSimilarity(GS + K_{best},SS)$}
\STATE $K_{best} \gets K_{i}$ 
\ENDIF
\ENDFOR
\IF{$cosSimilarity(GS + K_{best},SS) < cosSimilarity(GS,SS)$}
\RETURN $GS$ \{do not add $K_{best}$ if it worsens the similarity\}
\ENDIF
\ENDWHILE
\RETURN $GS$ 
\end{algorithmic}}
\end{algorithm}

The second algorithm is a greedy approach (Algorithm~\ref{alg:greedy}). It iteratively builds up a summary sentence-by-sentence: in each iteration, it determines the best-suited additional sentence and then adds it -- unless the addition of even the best-suited sentence would result in a worsening of the cosine similarity. 


\begin{algorithm}[t]
{\small\caption{Random Local Search with unrestricted summary length (RLS-unrestricted)}
\label{alg:RLS-unrestricted}
\KwInput{ 
    $AS$ - artefacts' sentences and $SS$ - student summary
}
\KwOutput{\textit{GS} – generated summary}
\begin{algorithmic}
\STATE $GS\gets $\O
\WHILE{(running time $< 10$ seconds)}
\STATE $GS_{temp} \gets GS$
\STATE select a sentence $AS_r$ from $AS$ u.a.r. and flip its inclusion status in $GS_{temp}$
\IF{$cosSimilarity(GS_{temp},SS) \geq cosSimilarity(GS,SS)$}
\STATE$GS \gets GS_{temp}$
\ENDIF
\ENDWHILE
\RETURN $GS$
\end{algorithmic}}
\end{algorithm}



In addition, we use three variations of random local search (RLS) algorithms. First, RLS-unrestricted (see Algorithms~\ref{alg:RLS-unrestricted}) can create summaries without being restricted by a target length. Second, RLS-restricted is like RLS-unrestricted, but it can only generate summaries of at most a given target length. Third, RLS-unrestricted-subset runs RLS-unrestricted first, but it then runs the brute force approach to find the best summary of at most a given target length. 
These algorithms share common characteristics, such as the execution time limit and the ability to explore the search space by including and excluding sentences. One notable characteristic of RLS-unrestricted is that it can produce summaries that exceed the target length. We have done this to provide an indication of whether five sentences were enough to create close summaries. 

As the sixth approach, we use a random search as a naive approach to provide a lower performance bound: it iteratively creates summaries of five sentences, and it returns (when the time is up) the best randomly created five-sentence summary. 

Note that the student summary ($SS$) that we provide as an input to all approaches can either be an actual summary (i.e., the words) in which case the co-occurrence is calculated, or it can be a summary represented as a feature vector in the high-dimensional feature space.

Lastly, to investigate the impact of the individual artefacts on the summaries, we consider three scenarios as input source to generate summaries by each of the algorithms at a given time window: 1) each of the artefacts listed in Table~\ref{tab:dataset} is considered individually as a source, 2) combining all the 15 artefacts in a single source, and 3) assuming we know a developer's preferences for particular types of artefacts, we only consider sentences coming from these types. 


\emph{Implementation note.} 
We remove a-posteriori all the cases when we encountered at least one empty summary for two reasons: (1) the word similarity between a generated summary and the student's summary can be zero, and (2) we encountered co-linear vectors even in the 26-dimensional space.

\section{Computational Study and Discussion}
\label{sec:experiments}

In our experiments, 
we consider the 503 summaries written by students, 6 algorithms, and three scenarios (i.e., the sentences' sources). 

For both similarity measures, we use the gold standard as the target, i.e., the students' original summaries either as bags of words or as high-dimensional feature vectors. An alternative for the feature similarity is to use, e.g., the average vector across all students to aim at the ``average style'', however, then it would not be clear anymore if it can be approximated. As this is the first such study, and in order to study the problem and the behaviour of the algorithms in this extractive setting under laboratory conditions, we aim for the solutions defined in the gold standard.

\paragraph{A comparison with brute force.}
To better understand what quality we can expect from our five randomised approaches, 
we compare these approaches with our brute force approach to extractive summarisation. The artefact type for this first investigation is ``issue title''. The maximum number of sentences here per project and summary combination was 35. For our brute force approach, this resulted in a manageable number of $324,632+52,360+6,545+595+35=384,167$ subsets of up to five sentences for that particular week. The computational budget that we give each RLS variant is 10 seconds.

\begin{figure}[t]
\vspace{-2mm}
\begin{subfigure}{.49\textwidth}
  \includegraphics[width=1\linewidth]{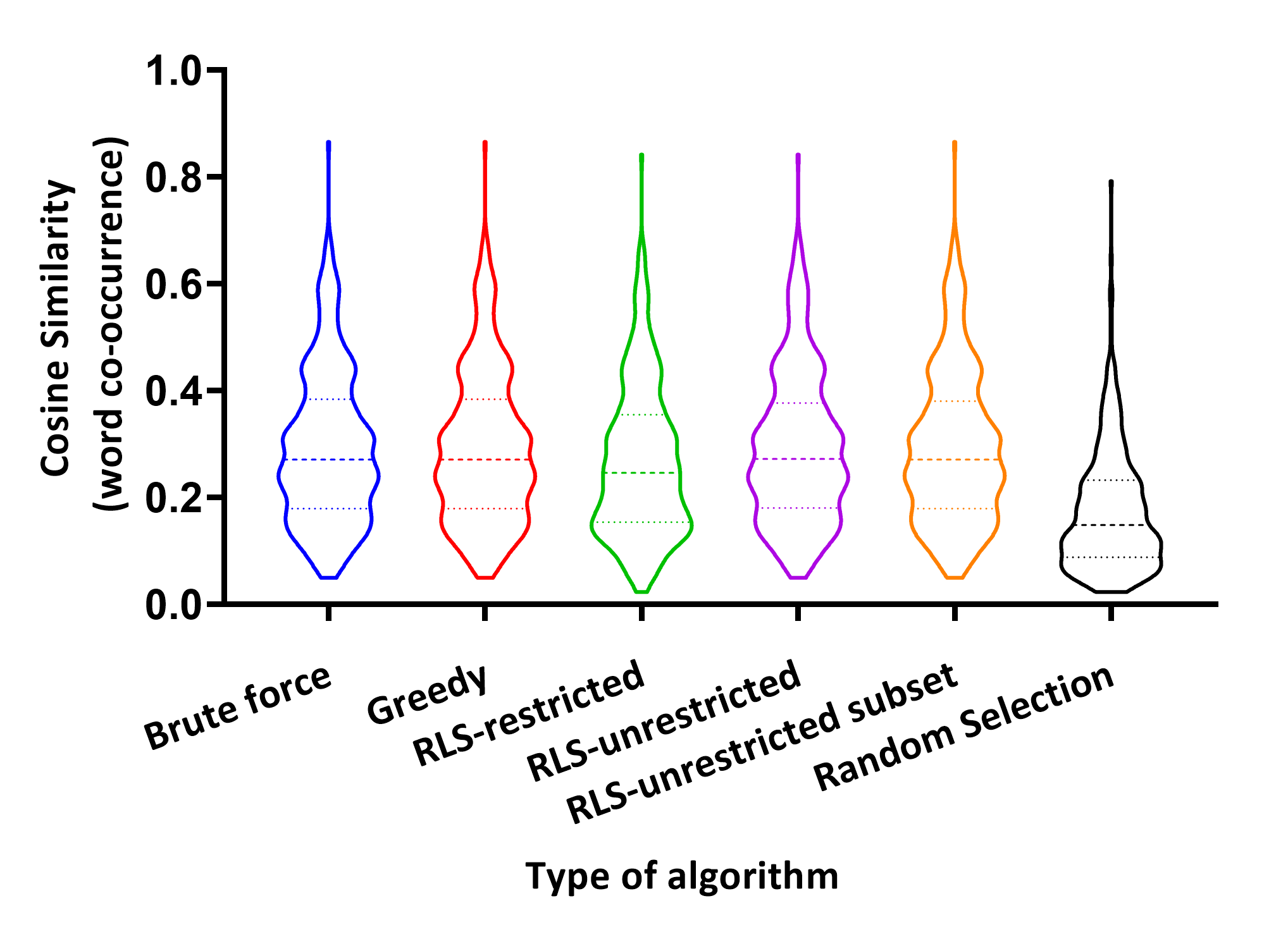}\vspace{-1mm}
  \caption{}
   \label{fig:bruteforceIssueTitle}
\end{subfigure}
\begin{subfigure}{.49\textwidth}
  \includegraphics[width=1\linewidth]{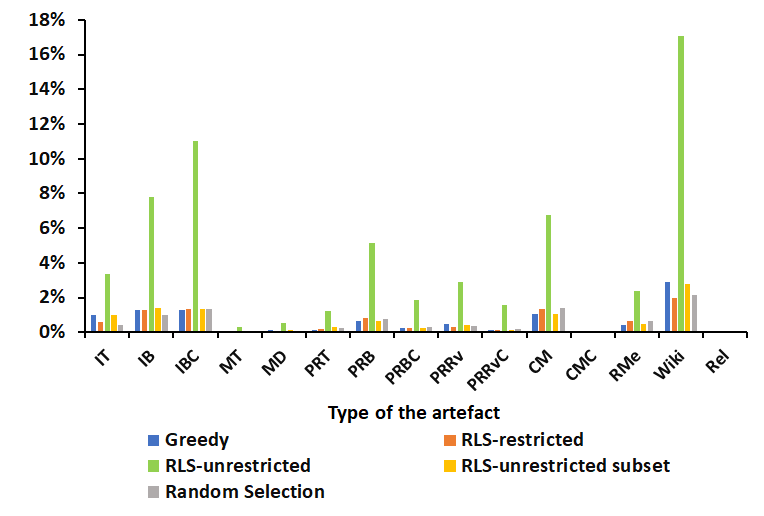}\vspace{-1mm}
  \caption{}
  \label{fig:TypeOfArtefactsByAlgorithmsAllArtefacts}
\end{subfigure}\\\vspace{-1mm}
\begin{subfigure}{.49\textwidth}
  \includegraphics[width=1\linewidth]{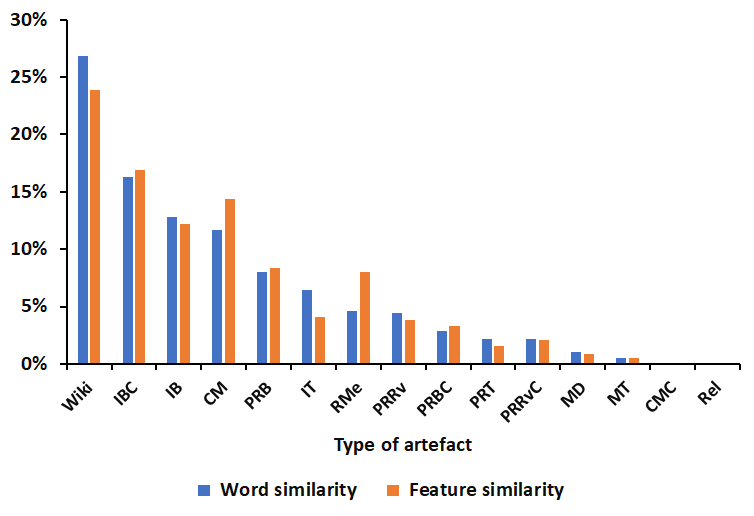}
  \caption{}
  \label{fig:TypeOfArtefactsAllArtefactsWordVsFeatyures}
\end{subfigure}
\centering
\begin{subfigure}{.49\textwidth}\vspace{-2mm}
  \includegraphics[width=1\linewidth]{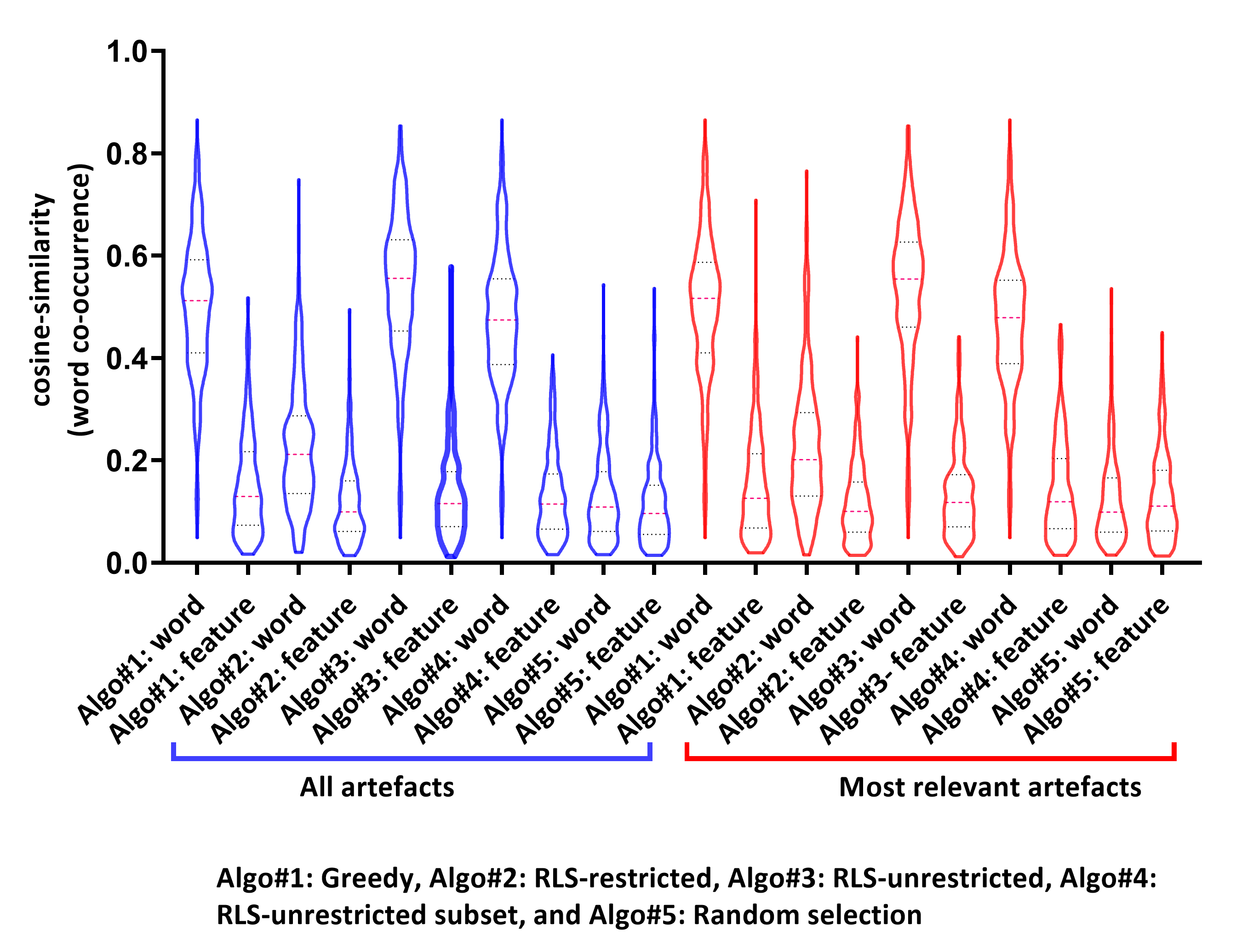}\vspace{-1mm}
  \caption{}
  \label{fig:AllMostArtefactsCosin}
\end{subfigure}%
\vspace{0mm}\caption{Results of the computational study. (a): Cosine similarity based on word co-occurrence of the generated summaries. 
(b): Average contribution of artefacts to summaries, aggregated across the two similarity measures. (c): Average contribution of artefacts to summaries, aggregated across all algorithms. (d): Similarities: when all artefacts are used (blue, overall average 0.266) and when only the relevant eight are used (red, overall average 0.258).}\vspace{-1mm}
\label{fig:eesults}
\end{figure}

Comparing the results obtained by these algorithms (see Figure~\ref{fig:bruteforceIssueTitle}), we can observe that the Greedy algorithm has the ability to generate summaries whose overall distribution is close to the distribution of summaries generated by brute force.\footnote{
Based on a two-sided Mann-Whitney U test, there is no statistically significant difference at p=0.05 between Greedy and Brute Force. 
}
Similarly, the two RLS-unrestricted approaches also produce comparable summaries. RLS-restricted performs worse, but still better than the Random Selection.\footnote{Let us recall let Random Selection does not generate only one summary at random, but many until the time limit is reached, and it then returns the best.} From this first comparison, we conclude that Greedy is a very good approach, as it achieves a performance comparable to that of brute force (which is our upper  performance bound), while it requires only 0.49 seconds on average compared to the 10 seconds of the RLS variants. 
We can moreover conclude that a maximum summary length of five is acceptable, as the \emph{RLS-unrestricted subset} does not perform differently from the others that were restricted.

To explain Greedy's performance, and to explain that the performances of Greedy and of some of the RLS variants is very comparable, we conjecture that the problem of maximising the cosine-similarity w.r.t. a target vector given a set of vectors is largely equivalent to a submodular 
pseudo-Boolean function without many local optima. A formal proof of this, however, 
remains future work. 




\paragraph{Used types of artefacts.} 

Next, use each algorithm to create a weekly summary for the cases where we have student summaries. In particular, we investigate from which artefact types the sentences are taken from in these generated summaries. 
In total, there are 22,313 (39.73\% of the total) sentences found in the source input linked to the students' summaries. Note that while this number appears to be very large, it includes the very large summaries produced by RLS-unrestricted (average length 29.6), and we are nevertheless aiming at hundreds of different target summaries for one-week time-windows, which thus appear to require very different sentences from the artefacts.

In Figure~\ref{fig:TypeOfArtefactsByAlgorithmsAllArtefacts}, we can see that the generated summaries by each of the algorithms are composed of sentences from almost all of the artefact types. In particular, we can note that sentences from wiki pages are most commonly used. Possible reasons for this include that (1) wiki pages make up the largest fraction of the source sentences, and (2) developers might have best described their activities on the wiki pages. 



In Figure~\ref{fig:TypeOfArtefactsAllArtefactsWordVsFeatyures}, we can see that content from wiki artefacts contributed around 27\% to the summaries generated by all algorithms. Also, sentences found in issue bodies (IB), issue body comments (IBC), and commit messages (CM) contributed 13\%--17\%. On the other hand, artefacts such as pull requests reviews (PRRv), pull requests title (PRT) and milestone titles (MT) have among the lowest contributions, which indicates that the students did not commonly use these artefacts -- or at least mention them and related content -- during their project's development life cycle. 


\paragraph{Generating summaries based on the most relevant artefacts.}

By generating summaries based on the most relevant artefacts found in the students' original summaries, we aim at generating more human-like summaries that better reflect the developers' preferences for certain artefact types. 
To achieve this 
we consider the generated summaries as a starting point, as each of them was generated to be similar to a particular student summary, and hence it can indirectly reflect a student's preference.  
Then, we identify the most relevant ones by using the median as the cut-off (i.e., based on Figure~\ref{fig:TypeOfArtefactsAllArtefactsWordVsFeatyures}). As a result of this selection, the eight most commonly referred to artefacts are (from most common to least common): wiki, issue title, issue bodies, issue body comments, commit messages, pull request bodies, readme files, and pull requests reviews. In total, this reduces the number of candidate sentences by 10.5\% to 50,246. 

We now investigate the performance of the subset of artefacts in terms of being able to generate good summaries.  
Figure~\ref{fig:AllMostArtefactsCosin} shows the cosine word co-occurrence similarity and feature similarity achieved by each of the algorithms. Blue violin plots show the distributions of similarities achieved when all 15 artefacts were considered, and the red violin pots show the same for the eight most relevant artefacts. 
As we can see, focusing on only eight artefact types appears to have little to no negative impact.


\section{Expert Annotation}
\label{sec:study}
\begin{table}[t]
\centering
\caption{Average rating from each annotator for output produced by the different approaches.}\vspace{-1mm}
\label{tab:annotators}
\begin{tabular}{lrr}
\toprule
\textbf{Approach} & \textbf{Annotator 1} & \textbf{Annotator 2}\\
\midrule
Word (all)         & 3.7   & 3.3      \\
Word (subset)      & \textbf{3.8}   & \textbf{3.5}      \\
Feature (all)      & 3.7   & 2.0      \\
Feature (subset)   & 3.7   & 2.0      \\
Random (all)       & 3.5   & 1.8      \\
Random (subset)    & 3.0   & 1.8      \\
\bottomrule
\end{tabular}
\end{table}

To evaluate the extent to which the summaries that the different approaches generate matched the summaries written by the students \emph{in the perception of software developers}, we asked two expert annotators to evaluate the results -- both were in their first year of study of a Computer Science PhD, and both not affiliated with this study. Both annotators indicated that developing software is part of their job, and they have 4--6 years of software development experience. Annotator 1 stated that they had 1--2 years of experience using GitHub for project development, Annotator 2 answered the same question with 2--4 years.

The selection of algorithms to be used for expert annotation is based on the highest median value of the cosine similarities between the gold standard summaries and the generated summaries from each of the algorithms. Therefore, summaries generated by the Greedy algorithm were chosen for the annotation. 

For the study, we randomly selected ten out of the total of fourteen weeks, and for each week, we randomly selected one project. For each of these ten, we then produced six different summaries in relation to the gold standard (i.e., the summaries written by the students):

\begin{enumerate}
    \item the best summary based on \textit{word similarity} between sentences contained in all artefacts in the input data (issues, pull requests, etc.) and the gold standard student summary,
    \item same as (1), but only using the eight most relevant artefacts as input data,
    \item the best summary based on \textit{feature similarity} between sentences contained in all artefacts in the input data and the gold standard student summary,
    \item same as (3), but only using the eight most relevant artefacts as input data,
    \item a random baseline by randomly selecting five sentences from all artefacts,
    \item same as (5), but only using the eight most relevant artefacts as input data.
\end{enumerate}

We created a questionnaire, 
which asked the annotators first to produce a summary for the ten selected weeks after inspecting the corresponding GitHub repositories (to ensure that annotators were familiar with the projects), and then to rate each summary on a Likert-scale from 1 (strongly disagree) to 5 (strongly agree) in response to the question ``Please indicate your agreement with the following statement: The summary mentions all important project activities present in the gold standard summary''.


Table~\ref{tab:annotators} shows the results, separately per annotator. While it is apparent from the data that Annotator 1 generally gave out higher scores than Annotator 2, both annotators perfectly agreed on the (partial) order of the different approaches: Word (subset) $\geq$ Word (all) $\geq$ Feature (subset) $\geq$ Feature (all) $\geq$ Random (all) $\geq$ Random (subset). 

In summary, approaches based on text similarity achieve the best result in terms of human perception, followed by approaches based on feature similarity, and the random baselines.



\section{Threats to Validity}
\label{sec:threats}
Our study, like many other studies, has a number of threats that may affect the validity of our results. 

First, our research subjects involved summaries written by graduate and undergraduate students. Although it is possible that Master students are more knowledgeable about interacting with the GitHub platform than the Bachelor students, the difference in the experiences of both subjects should not affect the results. This is because the students require an intermediate level of skills to work with the GitHub platform. 

Our result, illustrated in Table~\ref{tab:annotators},  
shows that the eight most relevant artefacts are found to be sufficient when generating summaries containing developers' activities. 
These types -- such as issues, pull requests, and commits -- are essential elements of a GitHub repository. However, as these are essential elements of probably any software repository, we expect this finding to be transferable to other repositories.

Also, evaluating the automatically generated summaries relies on human experts. Subjectivity and bias are likely to be issues when the number of human experts involved to assess the generated summaries is small. Hence, we plan, for future work, to include more experts to mitigate these issues. 


%
%
%

\section{Conclusion and Future Work}
\label{sec:conclusions}

Software engineering projects produce many artefacts over time, ranging from wiki pages, to pull request and issue comments. 
Summarising these can be helpful to a developer, for example, when a they return from a holiday, or when they try to get an overview of the project’s background in order to move forward with their team. 

In this article, we have presented a first framework to summarise the heterogeneous artefacts produced during a given time window. We have defined our own gold standard and ways of measuring similarity on a text-based level. Then, we proceeded to compare various optimisation heuristics and have found that a greedy approach performs best. A study then has found that experts preferred the combination that used word similarity to generate summaries based on the eight most relevant artefacts.

Interestingly, the generated summaries have been found useful even though the optimisation approaches have not yet considered temporal connections between the sentences and also not yet the actual meaning. In the next steps, we will focus on these two to further improve the quality of the summaries. An additional, larger study with Github users will aim at the use of averaged and personalised target vectors.



\subsection*{Acknowledgements}

Mahfouth has been sponsored by the Saudi Arabia Cultural Mission (SACM). 
Christoph's and Markus' work has been supported by the Australian Research Council projects DE180100153 and DE160100850.

\newpage

%
%
%
%
%
%


\end{document}